\documentclass[pra,aps,amsmath,amssymb,showkeys]{revtex4}
\newcommand{\pq}{\mathbf{p}}
\newcommand{\xp}{\mathbf{x}}
\newcommand{\zp}{\mathbf{z}}
\newcommand{\hh}{\mathcal{H}}
\newcommand{\lsa}{{\cal{L}}_{s.a.}}
\newcommand{\lsp}{{\cal{L}}_{+}}
\newcommand{\pen}{\openone}
\newcommand{\eqs}{E_{q}^{(s)}}
\newcommand{\rqm}{{\rm{E}}_{q}^{(s)}}
\newcommand{\tr}{{\rm{tr}}}
\newcommand{\wrho}{\widetilde{\rho}}

\newcommand{\ax}{\mathsf{X}}
\newcommand{\ay}{\mathsf{Y}}

\newcommand{\um}{\mathsf{U}}
\newcommand{\vm}{\mathsf{V}}
\newcommand{\mm}{\mathsf{M}}
\newcommand{\np}{\mathsf{P}}
\newcommand{\veps}{\varepsilon}
\newcommand{\rd}{{\rm{D}}}
\newcommand{\cp}{\mathcal{C}}
\newtheorem{Thm}{Theorem}[section]
\newtheorem{Lem}[Thm]{Lemma}
\begin{document}
\clearpage
\preprint{}

\title{Some general properties of unified entropies}
\author{Alexey E. Rastegin}
\affiliation{Department of Theoretical Physics, Irkutsk State University,
Gagarin Bv. 20, Irkutsk 664003, Russia}

\begin{abstract}
Basic properties of the unified entropies are examined. The
consideration is mainly restricted to the finite-dimensional
quantum case. Bounds in terms of ensembles of quantum states are
given. Both the continuity in Fannes' sense and stability in
Lesche's sense are shown for wide ranges of parameters. In particular,
uniform estimates are obtained for the quantum R\'{e}nyi entropies. Stability
properties in the thermodynamic limit are discussed as well. It is
shown that the unified entropies enjoy both the subadditivity and
triangle inequality for a certain range of parameters.
Non-decreasing of all the unified entropies under projective
measurements is proved.
\end{abstract}


\maketitle

\pagenumbering{arabic}
\setcounter{page}{1}

\section{Introduction}

The concept of entropy is one of fundamentals in statistical
physics and information theory. Entropic quantities are also
interesting mathematical subjects with many facets. Developing
statistical methods in various fields, some extensions of the
Shannon and von Neumann entropies were found to be expedient. Most
important of them are the R\'{e}nyi entropy \cite{renyi} and the
entropy introduced independently by Havrda and Charv\'{a}t
\cite{havrda} and by Tsallis \cite{tsallis}. Another class of
one-parameter extensions of the Shannon entropy was considered in
the paper \cite{arim71}. Hu and Ye proposed the concept of unified
entropy in both the classical and quantum regimes \cite{hey06}. Up
to a notation, this is actually the entropic form shortly
discussed by Tsallis in the review \cite{AO01}. All the above
entropies are particular cases of the unified entropies. At the
same time, some of so-called $f$-entropies \cite{czar08} are not
covered by the notion of unified entropy. Note that Petz's
quasi-entropies \cite{petz86,petz08} were emerged as the quantum
counterpart of the Csisz\'{a}r $f$-divergence.

There exist those general properties that are of primary
importance in physical applications of entropic functionals. In
the present paper, we analyze some of these properties with
respect to the family of unified entropies, mainly in the
finite-dimensional quantum regime. In particular, we consider
simple bounds in terms of pure-state ensembles and convex
combinations of density operators. The continuity in the sense of
Fannes as well as the stability in the sense of Lesche are
examined. Although the complete picture is not reached, few
important facts are resolved here. We prove the subadditivity of
the quantum unified entropy for a wide range of parameters. We
also show that all the quantum unified entropies are non-decreasing
under action of projective measurements.

The paper is organized as follows. In Section \ref{twbun}, the
main definitions are given. Some bounds in terms of ensembles of
quantum states are established. In Section \ref{unest}, uniform
estimates on the unified entropy for a wide range of parameter
values are obtained. These relations can be treated as
inequalities of Fannes' type. Using these results, in Section
\ref{nostb} we analyze the unified entropies with respect to the
Lesche stability criterion. Both the stability and instability in
the thermodynamic limit are considered as well. In Section
\ref{sbtin}, the known result on the subadditivity of quantum
Tsallis entropy for $q>1$ is naturally extended to some of the
unified entropies. In Section \ref{proen}, non-decreasing of the
quantum unified entropy under projective measurements is
established. It is also shown that measurements of more general
type can decrease the unified entropy. Section \ref{cocl}
concludes the paper.

\section{Definitions and two bounds}\label{twbun}

At first, we briefly recall some required material. Let $\lsa(\hh)$
be the space of Hermitian operators on $d$-dimensional Hilbert
space $\hh$. The set $\lsp(\hh)$ of positive semidefinite
operators is a convex cone in $\lsa(\hh)$. For any operator $\ax$,
we put $|\ax|\in\lsp(\hh)$ as the positive square root of
$\ax^{\dagger}\ax$. The singular values $\sigma_j(\ax)$ of
operator $\ax$ are then defined as the eigenvalues of $|\ax|$
\cite{bhatia97}. For $q\geq1$, the Schatten $q$-norm is defined by
\begin{equation}
\|\ax\|_{q}:=\left(\sum\nolimits_{j=1}^{d}\sigma_j(\ax)^q\right)^{1/q}
\ . \label{schndf}
\end{equation}
Both the R\'{e}nyi and Tsallis entropies are fruitful
one-parametric generalizations of the Shannon entropy. For unifying
treatment, we will denote the corresponding entropic index by $q$. Let
$\pq=\{p_1,\ldots,p_m\}$ be a probability distribution. For $q>0$
and $q\neq1$, the R\'{e}nyi $q$-entropy of this probability
distribution is defined as \cite{renyi}
\begin{equation}
R_{q}(\pq):=\frac{1}{1-q}{\ }\ln\left(\sum\nolimits_{i=1}^m p_i^{q}\right)
\ . \label{rendef}
\end{equation}
In the nonextensive statistical physics, the Tsallis $q$-entropy of the probability distribution $\pq$ is
defined as \cite{tsallis}
\begin{equation}
H_{q}(\pq):=\frac{1}{1-q}{\,}\left(\sum\nolimits_{i=1}^m p_i^{q} -1\right)
=-\sum\nolimits_{i=1}^m p_i^q\ln_q p_i
\ , \label{tsldef}
\end{equation}
where the $q$-logarithm $\ln_q x=\bigl(x^{1-q}-1\bigr)/(1-q)$.
Methods of nonextensive statistical mechanics have found use in
much many topics of physics and other sciences (see the works
\cite{GMT04,AO01} and references therein). In the limit $q\to1$,
both the above quantities coincide with the Shannon entropy
$H_{1}(\pq)=-\sum_ip_i\ln p_i$. Another extension is the entropy
of type $q$ \cite{arim71} defined as
\begin{equation}
{}_{q}H(\pq):=\frac{1}{q-1}{\,}\left[\left(\sum\nolimits_{i=1}^m p_i^{1/q}\right)^{q}-1 \right]
\ , \label{armdef}
\end{equation}
where $q>0$ and $q\neq1$. The authors of the paper
\cite{hey06} proposed the notion of unified entropy
\begin{equation}
E_{q}^{(s)}(\pq):=\frac{1}{(1-q){\,}s}{\,}\left[\left(\sum\nolimits_{i=1}^m p_i^{q}\right)^{{\!}s}-1 \right]
\label{unfdef}
\end{equation}
for $q>0$, $q\neq1$ and $s\neq0$. This two-parametric entropic
functional includes the entropies (\ref{tsldef}) and
(\ref{armdef}) as partial cases and the entropy (\ref{rendef}) as
the limiting case $s\to0$ \cite{hey06}. Taking $s=1-q'$, the
entropy (\ref{unfdef}) is actually the entropic $q,q'$-form
previously discussed by Tsallis \cite{AO01}. This form was emerged
in lines with the paper \cite{JR00}, where a nonextensive
thermodynamic formalism for description of chaos is developed.
Entropies of quantum states are important in various topics. The
quantum R\'{e}nyi entropy of density operator $\rho$ is defined as
\begin{equation}
{\rm{R}}_{q}(\rho):=\frac{1}{1-q}{\>}\ln\bigl[\tr(\rho^{q})\bigr]
\ , \label{qredf}
\end{equation}
that can be rewritten as
${\rm{R}}_{q}(\rho)=q(1-q)^{-1}\ln\|\rho\|_q$ for $q>1$. The
quantum Tsallis entropy of density operator $\rho$ is defined as
\begin{equation}
{\rm{H}}_{q}(\rho):=\frac{1}{1-q}{\>}\tr\bigl(\rho^{q}-\rho\bigr)
\ , \label{qtsdf}
\end{equation}
in particular,
${\rm{H}}_{q}(\rho)=(1-q)^{-1}\left(\|\rho\|_q^q-1\right)$ for
$q>1$. For $q=1$, we have the von Neumann entropy
${\rm{S}}(\rho)=-\tr(\rho\ln\rho)$. Like the R\'{e}nyi and Tsallis
entropies, the expression (\ref{unfdef}) can be adopted for the
quantum case. The quantum unified $(q,s)$-entropy of density
operator $\rho$ is defined as
\begin{equation}
\rqm(\rho):=\frac{1}{(1-q){\,}s}{\>}
\Bigl\{\bigl[\tr(\rho^{q})\bigr]^s-1\Bigr\}
\label{qundef}
\end{equation}
for $q>0$, $q\neq1$ and $s\neq0$, and as the quantum R\'{e}nyi
entropy ${\rm{E}}_{q}^{(0)}(\rho)={\rm{R}}_{q}(\rho)$ for $s=0$
\cite{hey06}. Some properties of the quantum unified entropy were
considered in \cite{hey06}. In particular, lower and upper bounds
on the entropy (\ref{qundef}) are given for $1\leq{q}$,
$1\leq{s}$. We now obtain two results which complement the
discussion of the paper \cite{hey06} in this regard.

\begin{Thm}\label{conun}
Let $\bigl\{p_i,|\psi_i\rangle\bigr\}$, where
$\langle\psi_i|\psi_i\rangle=1$ and $\sum_ip_i=1$, be an ensemble
of pure states that gives rise to the density operator $\rho$,
i.e.
\begin{equation}
\rho=\sum\nolimits_i p_i{\,}|\psi_i\rangle\langle\psi_i|
\ . \label{ensres}
\end{equation}
For $0<q$ and $s\neq0$ as well as for $0<q<1$ and $s=0$, there holds
\begin{equation}
\rqm(\rho)\leq\eqs(\pq)
\ . \label{cnuneq}
\end{equation}
\end{Thm}

{\bf Proof.} We will assume that $q\neq1$. Let $\lambda_j$ and
$|\varphi_j\rangle$ denote eigenvalues and eigenstates of
$\rho=\sum_j\lambda_j{\,}|\varphi_j\rangle\langle\varphi_j|$. The
ensemble classification theorem says that \cite{hugston}
\begin{equation}
\sqrt{p_i}{\,}|\psi_i\rangle=\sum\nolimits_j u_{ij}\sqrt{\lambda_j}{\,}|\varphi_j\rangle
\label{unfr}
\end{equation}
for some unitary matrix $[[u_{ij}]]$. It follows from (\ref{unfr})
and $\langle\varphi_j|\varphi_k\rangle=\delta_{jk}$ that
$p_i=\sum_jw_{ij}{\,}\lambda_j$, where
$w_{ij}=u_{ij}^{*}{\,}u_{ij}$ are elements of a unistochastic
matrix, i.e. $\sum\nolimits_{i}w_{ij}=1$ for all $j$ and
$\sum\nolimits_{j}w_{ij}=1$ for all $i$. The function $x\mapsto
x^{q}$ is concave for $0<q<1$ and convex for $1<q$. Applying
Jensen's inequality to this function, we obtain
\begin{equation}
\sum\nolimits_i p_i^{q}=
\sum\nolimits_i\left(\sum\nolimits_jw_{ij}{\,}\lambda_j\right)^{q}
{\ }\left\{
\begin{array}{cc}
\geq, & 0<q<1 \\
\leq, & 1<q
\end{array}
\right\}
{\ }\sum\nolimits_i \sum\nolimits_jw_{ij}{\,}\lambda_j^{q}
=\sum\nolimits_j \lambda_j^{q}
\label{agal1}
\end{equation}
in view of unistochasticity of the matrix $[[w_{ij}]]$. The
function $y\mapsto y^s/s$ monotonically increases for $s\neq0$,
whence
\begin{equation}
\frac{1}{s}{\>}\left(\sum\nolimits_i p_i^{q}\right)^{{\!}s}
{\ }\left\{
\begin{array}{cc}
\geq, & 0<q<1 \\
\leq, & 1<q
\end{array}
\right\} {\ }\frac{1}{s}{\>}\left(\sum\nolimits_j
\lambda_j^{q}\right)^{{\!}s} \ . \label{agal2}
\end{equation}
Since the factor $(1-q)^{-1}$ is positive for $q<1$ and negative
for $1<q$, the relations (\ref{agal2}) are combined as
\begin{equation}
\frac{1}{(1-q){\,}s}{\>}\left(\sum\nolimits_i p_i^{q}\right)^{{\!}s}
\geq\frac{1}{(1-q){\,}s}{\>}\left(\sum\nolimits_j \lambda_j^{q}\right)^{{\!}s}
\ . \label{agal3}
\end{equation}
Due to $\sum_j\lambda_j^{q}=\tr(\rho^{q})$ and the definitions
(\ref{unfdef}) and (\ref{qundef}), the inequality (\ref{agal3})
provides (\ref{cnuneq}). In the case $s=0$, when the quantum
R\'{e}nyi entropy (\ref{qredf}) is dealt, the result
(\ref{cnuneq}) for $0<q<1$ has been proved in the paper
\cite{rast104}. $\blacksquare$

The statement of Theorem \ref{conun} gives an upper bound on the
quantum unified entropy in terms of ensembles of pure states. For
$q=1$, when the von Neumann entropy is dealt, the property
(\ref{cnuneq}) is well known \cite{wehrl}. We have seen that the
same holds for almost all of the unified entropies. Another upper
bound in terms of ensembles of mixed states has been proved in the
paper \cite{hey06}, but only for the parameter range
$\bigl\{(q,s):{\>}1\leq{q},{\>}1\leq s\bigr\}$. So the scope of
the inequality (\ref{cnuneq}) is much more wide in both the
parameters $q$ and $s$. The second result is related to convex
combination of density operators. Suppose that
\begin{equation}
\rho=\sum\nolimits_{i}p_i{\,}\omega_i
\ , \label{hrom}
\end{equation}
where $\tr(\omega_i)=1$ and $\sum_ip_i=1$. For $1\leq{q}$ and
$1\leq{s}$, the authors of the paper \cite{hey06} also presented a
lower bound on $\rqm(\rho)$ in terms of combinations of a kind
(\ref{hrom}). So we shall now discuss this inequality for the
range $0<q<1$.

\begin{Thm}\label{conoc}
For $0<q<1$, $s\leq1$ and any convex combination (\ref{hrom}),
there holds
\begin{equation}
\sum\nolimits_{i}p_i{\,}\rqm(\omega_i)\leq\rqm(\rho)
\ . \label{cnceq1}
\end{equation}
\end{Thm}

{\bf Proof.} We firstly assume that $s\neq0$. For any concave
function $f(x)$, the functional $\tr\bigl[f(\ax)\bigr]$ is also
concave on $\ax\in\lsa(\hh)$ (for details, see section III in
\cite{rast10fn}). For $0<q<1$, the function $x\mapsto{x}^q$ is
just concave, whence
\begin{equation}
\tr(\rho^{q})\geq\sum\nolimits_{i}p_i{\,}\tr(\omega_i^{q})
\ . \label{romi}
\end{equation}
The function $y\mapsto{y}^s/s$ on $y\in[0;\infty)$ is increasing
for all $s\neq0$ and concave for $s\leq1$ (its second derivative
$(s-1){\,}y^{s-2}\leq0$). Applying these two points step-by-step,
we have
\begin{equation}
\frac{1}{s}{\>}\bigl[\tr(\rho^{q})\bigr]^s\geq
\frac{1}{s}{\,}\left[\sum\nolimits_ip_i{\,}\tr(\omega_i^{q})\right]^s
\geq\frac{1}{s}{\,}\sum\nolimits_ip_i{\,}\bigl[\tr(\omega_i^{q})\bigr]^s
\label{cass}
\end{equation}
for $s\leq1$. Multiplying the term (\ref{cass}) by $(1-q)^{-1}>0$
and using the definition (\ref{qundef}), we obtain the claim
(\ref{cnceq1}). When $s=0$, the quantum R\'{e}nyi entropy
(\ref{qredf}) is dealt. So we merely rewrite the relations
(\ref{cass}) with the function $y\mapsto\ln{y}$, which is also
increasing and concave. $\blacksquare$

In the parameter range
$\bigl\{(q,s):{\>}1\leq{q},{\>}1\leq{s}\bigr\}$, the bound
(\ref{cnceq1}) has been given in the paper \cite{hey06}. The
statement of Theorem \ref{conoc} implies the concavity property of
the quantum unified entropy for $0<q<1$ and $s\leq1$. The
corresponding property of the von Neumann entropy is well known
\cite{wehrl}. It must be stressed that the R\'{e}nyi entropy of
order $q>1$ is not purely convex nor purely concave \cite{jiar04},
even in the classical regime. In general, the inequality
(\ref{cnceq1}) can also be regarded as lower bounds on the quantum
unified entropy in terms of convex combination of density
operators. Thus, we can summarize the following. In a wide range
of parameter values, the quantum unified entropy (\ref{qundef})
enjoy concavity properties similarly to the von Neumann entropy.
Since such properties are very important for entropic functionals,
potential capabilities of the unified entropies become more clear with the
above results.

\section{Uniform estimates}\label{unest}

Any entropic quantity must have good properties as a state
functional. First of all, we are interested in continuity. For the
von Neumann entropy, the answer is given by the well-known Fannes
inequality \cite{fannes}. This inequality in terms of trace norm
distance bounds from above a potential change of the von Neumann
entropy. Fannes' inequality has been extended to the Tsallis
entropy \cite{yanagi,zhang} and its partial sums \cite{rast1023}.
Some bounds of Fannes' type on the quantum relative entropy were
obtained in the paper \cite{AE05}. As given in terms of difference
distances, these bounds characterize a continuity of the quantum
relative entropy in the sense of Fannes. We shall examine the
unified entropy from this viewpoint.

Recall some results related to the Tsallis entropy. In the
considered question, both the classical and quantum regimes can be
treated simultaneously. So we will formulate for the quantum one.
The trace distance between two operators $\ax$ and $\ay$ is
defined as
\begin{equation}
\rd_{\tr}(\ax,\ay):=\frac{1}{2}{\>}\tr|\ax-\ay|=\frac{1}{2}{\>}\|\ax-\ay\|_{1}
\ . \label{trdf}
\end{equation}
In the classical regime, the trace distance is replaced with
the $l_1$-distance. The two complementary results are known for
the Tsallis entropy. The first method of estimating with
restriction $q\in[0;2]$ is presented in the paper \cite{yanagi}.
Let us denote $\eta_{q}(x):=\bigl(x^{q}-x\bigr)/(1-q)$.
For $q\in[0;2]$ and $d={\rm{dim}}(\hh)$, we have
\begin{equation}
\bigl|{\rm{H}}_{q}(\rho)-{\rm{H}}_{q}(\omega)\bigr|\leq(2\veps)^{q}\ln_{q}d+\eta_{q}(2\veps)
\label{resfk}
\end{equation}
provided that $2\rd_{\tr}(\rho,\omega)=2\veps\leq{q}^{1/(1-q)}$.
On the interval $0<2\veps<{q}^{1/(1-q)}$, the quantity
$\eta_{q}(2\veps)$ and hence the right-hand side of (\ref{resfk})
monotonically increase. The second method uses probabilistic
coupling techniques \cite{zhang}. For $q>1$, we then have
\begin{equation}
\bigl|{\rm{H}}_{q}(\rho)-{\rm{H}}_{q}(\omega)\bigr|\leq\veps^{q}\ln_{q}(d-1)+
H_{q}(\veps,1-\veps)
\ , \label{reszh}
\end{equation}
where $\veps=\rd_{\tr}(\rho,\omega)$ and
$H_{q}(\veps,1-\veps)=(1-q)^{-1}\bigl(\veps^{q}+(1-\veps)^{q}-1\bigr)$
denotes the binary entropy. By calculus, the right-hand side of
(\ref{reszh}) monotonically increases for $0<\veps<(d-1)/d$. The
inequality (\ref{reszh}) is sharp in the sense that the equality
sometimes takes place \cite{zhang}. Including the dimension $d$
explicitly, the above bounds are finite-dimensional. The upper bounds (\ref{resfk}) and
(\ref{reszh}) can be used for obtaining upper bounds on the
unified entropy for some parameter ranges. Our method is based
on the following statement.

\begin{Lem}\label{sist}
For $s\geq1$ and $x,y\in[0;1]$ as well as for $0<s\leq1$ and
$x,y\in[1;\infty)$, there holds
\begin{equation}
|x^s-y^s|\leq
s{\,}|x-y|
\ . \label{xssy}
\end{equation}
For $0<s\leq1$ and $x,y\in[0;1]$ as well as for $s\geq1$ and
$x,y\in[1;\infty)$, the reversed inequality holds. For
$x,y\in[1;\infty)$, the following inequality also takes place:
\begin{equation}
|\ln{x}-\ln{y}|\leq
|x-y|
\ . \label{lnsy}
\end{equation}
\end{Lem}

{\bf Proof.} When $x<y$, the aim (\ref{xssy}) and the reversed one
are recast as $(y^s-x^s)\leq{s}{\,}(y-x)$ and
$(y^s-x^s)\geq{s}{\,}(y-x)$ respectively. Here we write down
\begin{equation}
y^s-x^s=\int\nolimits_{x}^{y}s{\,}t^{s-1}dt
{\ }\left\{
\begin{array}{cc}
\leq, & t^{s-1}\leq1 \\
\geq, & t^{s-1}\geq1
\end{array}
\right\}
{\ }
\int\nolimits_{x}^{y}s{\,}dt=s{\,}(y-x)
\ . \label{xyin}
\end{equation}
The condition $t^{s-1}\leq1$ is satisfied for $0<s<1$ and $t\geq1$
as well as for $s>1$ and $t\leq1$. The condition $t^{s-1}\geq1$ is
satisfied for $0<s<1$ and $t\leq1$ as well as for $s>1$ and
$t\geq1$. The aim (\ref{lnsy}) is justified by
\begin{equation}
\ln{y}-\ln{x}=\int\nolimits_{x}^{y}\frac{dt}{t}\leq\int\nolimits_{x}^{y}dt=y-x
\end{equation}
under the condition $t\geq1$. $\blacksquare$

Note that the case $s=1$ is the scope of the continuity bounds
(\ref{resfk}) and (\ref{reszh}) for the Tsallis entropy. We do not
take $q=1$, since the von Neumann entropy itself is considered in
the original Fannes inequality \cite{fannes}. The main result of
this section is formulated as follows.

\begin{Thm}
Let $\rho$ and $\omega$ be density operators on $d$-dimensional
Hilbert space $\hh$. For the parameter range
\begin{equation}
\bigl\{(q,s):{\>}0<q<1,{\>}s\in(-\infty;-1]\cup[0;+1]\bigr\}
\label{pdm1}
\end{equation}
under the condition
$2\rd_{\tr}(\rho,\omega)=2\veps\leq{q}^{1/(1-q)}$, there holds
\begin{equation}
\bigl|{\rm{E}}_{q}^{(s)}(\rho)-{\rm{E}}_{q}^{(s)}(\omega)\bigr|
\leq(2\veps)^{q}\ln_{q}d+\eta_{q}(2\veps)
\ . \label{resfk1}
\end{equation}
Define the factor $\varkappa_s=d^{2(q-1)}$ for $s\in[-1;0]$ and
$\varkappa_s=1$ for $s\in[+1;+\infty)$. For the parameter range
\begin{equation}
\bigl\{(q,s):{\>}1<q,{\>}s\in[-1;0]\cup[+1;+\infty)\bigr\}
\label{pdp1}
\end{equation}
and arbitrary $\rd_{\tr}(\rho,\omega)=\veps$, there holds
\begin{equation}
\bigl|{\rm{E}}_{q}^{(s)}(\rho)-{\rm{E}}_{q}^{(s)}(\omega)\bigr|
\leq\varkappa_s\bigl[\veps^{q}\ln_{q}(d-1)+H_{q}(\veps,1-\veps)\bigr]
\ . \label{reszh1}
\end{equation}
\end{Thm}

{\bf Proof.} Let us put the quantities $\xi=\tr(\rho^{q})$ and
$\zeta=\tr(\omega^{q})$. For the range (\ref{pdm1}), we have
$\xi\geq1$ and $\zeta\geq1$ due to $0<q<1$. For
$s\in(-\infty;-1]$, we put $\nu=-s\geq+1$ and then write
\begin{equation}
\bigl|{\rm{E}}_{q}^{(s)}(\rho)-{\rm{E}}_{q}^{(s)}(\omega)\bigr|=
\frac{1}{(1-q)\nu}{\,}\left|\frac{1}{\xi^{\nu}}-\frac{1}{\zeta^{\nu}}\right|
\ . \label{mem1}
\end{equation}
Using the inequality (\ref{xssy}) with $x=1/\xi$, $y=1/\zeta$ and
$\nu$ instead of $s$, the formula (\ref{mem1}) leads to
\begin{equation}
\bigl|{\rm{E}}_{q}^{(s)}(\rho)-{\rm{E}}_{q}^{(s)}(\omega)\bigr|\leq
\frac{1}{(1-q)}{\ }\frac{|\zeta-\xi|}{\xi\zeta}
\leq\frac{1}{1-q}{\>}|\xi-\zeta|=\bigl|{\rm{H}}_{q}(\rho)-{\rm{H}}_{q}(\omega)\bigr|
\label{mem2}
\end{equation}
due to the definition (\ref{qtsdf}). Putting $x=\xi$ and
$y=\zeta$, we use the inequalities (\ref{xssy}) for $s\in(0;+1]$
and (\ref{lnsy}) for $s=0$ (the R\'{e}nyi entropy case) with the
result
\begin{equation}
\bigl|{\rm{E}}_{q}^{(s)}(\rho)-{\rm{E}}_{q}^{(s)}(\omega)\bigr|
\leq\left|\frac{\xi-\zeta}{1-q}\right|=\bigl|{\rm{H}}_{q}(\rho)-{\rm{H}}_{q}(\omega)\bigr|
\ . \label{pem2}
\end{equation}
Combining the right-hand side of (\ref{mem2}) and (\ref{pem2})
with the bound (\ref{resfk}) finally gives (\ref{resfk1}).

For the range (\ref{pdp1}), we have $\xi\leq1$ and $\zeta\leq1$ by
$q>1$. In addition, both the traces $\xi=\tr(\rho^{q})$ and
$\zeta=\tr(\omega^{q})$ are not less than $d^{1-q}$. For
$s\in[-1;0)$, we take $\nu=-s\in(0;+1]$ and write down
\begin{equation}
\bigl|{\rm{E}}_{q}^{(s)}(\rho)-{\rm{E}}_{q}^{(s)}(\omega)\bigr|=
\frac{1}{(q-1)\nu}{\,}\left|\frac{1}{\xi^{\nu}}-\frac{1}{\zeta^{\nu}}\right|
\leq\frac{1}{(q-1)}{\ }\frac{|\zeta-\xi|}{\xi\zeta}
\leq\frac{d^{2(q-1)}}{q-1}{\>}|\xi-\zeta|=d^{2(q-1)}\bigl|{\rm{H}}_{q}(\rho)-{\rm{H}}_{q}(\omega)\bigr|
\label{mem0}
\end{equation}
by (\ref{xssy}) with $x=1/\xi$, $y=1/\zeta$ and $\nu$ instead of
$s$. When $s=0$ (the R\'{e}nyi entropy case), we use (\ref{lnsy})
and get the upper bound (\ref{mem0}) as well. For $+1\leq{s}$, we
merely apply the formula (\ref{xssy}) with $x=\xi$, $y=\zeta$ and
obtain (\ref{pem2}) again. Combining the relations (\ref{pem2})
and (\ref{mem0}) with (\ref{reszh}) finally gives (\ref{reszh1}).
$\blacksquare$

For $1\leq{q}$ and $+1\leq{s}$, the authors of the paper
\cite{hey06} have presented the upper bound
\begin{equation}
\bigl|{\rm{E}}_{q}^{(s)}(\rho)-{\rm{E}}_{q}^{(s)}(\omega)\bigr|
\leq2q(q-1)^{-1}\rd_{\tr}(\rho,\omega)
\ . \label{lipb}
\end{equation}
So the inequality (\ref{reszh1}) gives another bound for such
values of $q$ and $s$. But the novel point is that the bound
(\ref{reszh1}) is also valid for values $1\leq{q}$ and
$s\in[-1;0]$. In addition, for the parameter range (\ref{pdm1})
with $0<q<1$, the upper bound is stated by the formula
(\ref{resfk1}). Like the bounds (\ref{resfk}) and (\ref{reszh}),
the new bounds (\ref{resfk1}) and (\ref{reszh1}) are related to
finite dimensions solely. These bounds are also not sharp.
Nevertheless, they show that the unified entropy is continuous in
the sense of Fannes for the parameter ranges (\ref{pdm1}) and
(\ref{pdp1}). Moreover, the above results lead to some conclusions
on the stability.

\section{Notes on stability properties}\label{nostb}

The stability property is very important for application of some
quantity in a practice. This issue was inspired by Lesche
\cite{lesche} who discussed the Renyi entropy. The Tsallis entropy
itself \cite{abe02} and its partial sums \cite{rast1023} were
considered with respect to the stability property. The class of
$f$-entropies is examined in the papers \cite{CL09,naudts04}. In
this section, we analyze the unified entropies from this viewpoint.
The Lesche stability criterion can be posed as follows
\cite{abe02,CL09}. Let $\Phi(\rho)$ be a state
functional. We assume its values to be bounded from above by the least
bound $\Phi_M$. The stability means that for every $\delta>0$,
there exists $\varepsilon>0$ such that
\begin{equation}
|\Phi({\rho})-\Phi({\omega})|\cdot\Phi_M^{-1}\leq\delta
\ , \label{lscdf}
\end{equation}
whenever $\rd_{\tr}(\rho,\omega)\leq\veps$. In the
finite-dimensional case, the unified entropy is bounded from above
for all real $s$ and $q>0$ \cite{hey06}. In particular, for
$q\neq1$ and $s\neq0$ we have
\begin{equation}
\rqm(\rho)\leq\max\rqm=\frac{1}{(1-q){\,}s}{\,}\left(d^{(1-q)s}-1\right)
\ , \label{hyub}
\end{equation}
and $\max{\rm{E}}_q^{(0)}=\ln{d}$. Using the bounds (\ref{resfk1})
and (\ref{reszh1}), for the parameter ranges (\ref{pdm1}) and
(\ref{pdp1}) we obtain
\begin{equation}
\bigl|{\rm{E}}_{q}^{(s)}(\rho)-{\rm{E}}_{q}^{(s)}(\omega)\bigr|\cdot\bigl(\max\rqm\bigr)^{-1}\leq{F}(\veps)
\ . \label{emef}
\end{equation}
Here the function $F(\veps)$ respectively denotes the right-hand
sides of (\ref{resfk1}) and (\ref{reszh1}) multiplied by
$\bigl(\max\rqm\bigr)^{-1}$. This function obeys $F(0)=0$ and
monotonically increases with $\veps\in(0;\veps_0)$, where
$2\veps_0={q}^{1/(1-q)}$ for (\ref{pdm1}), $\veps_0=(d-1)/d$ for
(\ref{pdp1}). Hence one defines some one-to-one correspondence
between the intervals $[0;\veps_0]$ and $[0;\delta(\veps_0)]$. For
each $\delta\in(0;\delta(\veps_0)]$ we herewith have
$\veps_{\delta}>0$ such that $F(\veps)\leq\delta$ whenever
$\veps\leq\veps_{\delta}$. Combining this with the relation
(\ref{emef}) implies the stability property of the unified entropy
for the parameter ranges (\ref{pdm1}) and (\ref{pdp1}).

The case $s=0$, when R\'{e}nyi's entropy is dealt, demands few
remarks. It is a common opinion that the R\'{e}nyi entropy of
order $q$ is not stable for all $q\neq1$
\cite{abe02,CL09,naudts04}. This important conclusion was obtained
by Lesche \cite{lesche} and developed by Abe \cite{abe02}.
Meantime, both the parameter ranges (\ref{pdm1}) and (\ref{pdp1})
include the R\'{e}nyi entropy. No contradiction is really posed
here. Indeed, the reasons of the papers \cite{abe02,lesche} assume
the thermodynamic limit, i.e. an infinity of degrees of freedom.
But the above bounds are just finite-dimensional. Dividing
(\ref{resfk1}) and (\ref{reszh1}) by
$\max{\rm{E}}_q^{(0)}=\ln{d}$, the right-hand side of (\ref{emef})
contains the ratio
$\ln_{q}d/\ln{d}\approx(1-q)^{-1}{d}^{1-q}/\ln{d}$ for $0<q<1$,
$s=0$, and the one $d^{2(q-1)}/\ln{d}$ for $1<q$, $s=0$. They are
both unbounded as $d\to\infty$. In this limit, no conclusions on
the R\'{e}nyi entropy can be got from the inequality (\ref{emef}).
The stability of the R\'{e}nyi entropy of order $q\neq1$ does hold
for a finite $d$, but does not in the thermodynamic limit. In a
contrast, the Tsallis entropy enjoys the following. When $0<q<1$,
we have (\ref{resfk}) and $F(\veps)\to(2\veps)^{q}$ in the limit
$d\to\infty$; when $q>1$, we have (\ref{reszh}) and
$F(\veps)\to\veps^{q}+(q-1){\,}H_{q}(\veps,1-\veps)$.

Since the thermodynamic limit is of great importance in
statistical physics, we shall now discuss this issue for the
unified entropies. We divide the relations (\ref{resfk1}) and
(\ref{reszh1}) by $\max\rqm$ and take the limit $d\to\infty$. Some
inspection shows that the stability in the thermodynamic limit
holds for the case  $1\leq{q}$ and $+1\leq{s}$. Taking $q\neq1$
and combining (\ref{reszh1}) and (\ref{hyub}), for this case we
have
\begin{equation}
\frac{\bigl|{\rm{E}}_{q}^{(s)}(\rho)-{\rm{E}}_{q}^{(s)}(\omega)\bigr|}{\max\rqm}\leq
\frac{(q-1)s}{1-d^{(1-q)s}}\left[\veps^q{\>}\frac{(d-1)^{1-q}-1}{1-q}+H_{q}(\veps,1-\veps)\right]
\underset{d\to\infty}{\longrightarrow}
s\veps^q+(q-1)sH_{q}(\veps,1-\veps)
\ . \label{sqg1}
\end{equation}
Rewriting the right-hand side of (\ref{sqg1}) as
$s\bigl[1-(1-\veps)^q\bigr]\leq{s}q\veps$ by (\ref{xssy}), the
stability property is obvious from this relation. For remaining
parts of the ranges (\ref{pdm1}) and (\ref{pdp1}), the right-hand
side of (\ref{emef}) is not bounded as $d\to\infty$. So, no conclusions can be
made with the inequality (\ref{emef}) for these parts.

Instabilities are usually observed by constructing an explicit
example. We will adopt the two examples employed in the paper
\cite{lesche}. Consider the two commuting density operators
defined as
\begin{equation}
\rho_{0}={\rm{diag}}\bigl(1,0,\ldots,0\bigr)
\ , \qquad \omega_{0}={\rm{diag}}\left(1-\veps,\frac{\veps}{d-1},\ldots,\frac{\veps}{d-1}\right),
\label{roso}
\end{equation}
in their eigenbasis. It is clear that
$\rd_{\tr}(\rho_0,\omega_0)=\veps$. Further, for all $q>0$ we have
$\tr(\rho_0^q)=1$ and
\begin{equation}
\tr(\omega_0^q)=(1-\veps)^q+\veps^q(d-1)^{1-q}
\ . \label{sigq}
\end{equation}
Note that the inequality (\ref{reszh}) is saturated with the
states (\ref{roso}). We claim that the unified entropy is not
stable in the thermodynamic limit for $0<q<1$ and $s<0$. Indeed,
we have $(1-q)s<0$ and
\begin{equation}
\frac{\bigl|{\rm{E}}_{q}^{(s)}(\rho_0)-{\rm{E}}_{q}^{(s)}(\omega_0)\bigr|}{\max\rqm}
=\frac{\left|1-\bigl[\tr(\omega_0^q)\bigr]^s\right|}{1-d^{(1-q)s}}
\underset{d\to\infty}{\longrightarrow}1
\label{sqg11}
\end{equation}
in view of $\tr(\omega_0^q)\to\infty$ and $s<0$. Hence the
stability condition is clearly violated.

Consider now the two commuting density operators defined as
\begin{equation}
\rho_{1}={\rm{diag}}\left(0,\frac{1}{d-1},\ldots,\frac{1}{d-1}\right),
\qquad \omega_{1}={\rm{diag}}\left(\veps,\frac{1-\veps}{d-1},\ldots,\frac{1-\veps}{d-1}\right),
\label{roso1}
\end{equation}
where $\rd_{\tr}(\rho_1,\omega_1)=\veps$ as well. Here we get
$\tr(\rho_1^q)=(d-1)^{1-q}$ and
$\tr(\omega_1^q)=\veps^q+(1-\veps)^q(d-1)^{1-q}$. This example
allows to resolve a violation of the stability for $1<q$ and
$s<0$. For $d\gg1$, we can asymptotically take
\begin{equation}
\max\rqm\approx\frac{1}{(1-q){\,}s}{\>}d^{(1-q)s}
\ , \qquad
\tr(\omega_1^q)\approx\veps^q
\ , \label{hyub1}
\end{equation}
in view of $(1-q)s>0$ and $(1-q)<0$ respectively. Due to the
relations (\ref{hyub1}), there holds
\begin{equation}
\frac{\bigl|{\rm{E}}_{q}^{(s)}(\rho_1)-{\rm{E}}_{q}^{(s)}(\omega_1)\bigr|}{\max\rqm}\approx
\frac{\bigl|(d-1)^{(1-q)s}-\veps^{qs}\bigr|}{d^{(1-q)s}}
\underset{d\to\infty}{\longrightarrow}1
\ . \label{sqg111}
\end{equation}
Thus, the stability condition is also violated in the
thermodynamic limit for $1<q$ and $s<0$.

The above results can be summarized as follows. For $1\leq{q}$ and
$+1\leq{s}$, the unified entropy is actually stable including the
thermodynamic limit. For other parts of the parameter ranges
(\ref{pdm1}) and (\ref{pdp1}), the stability in the
finite-dimensional case holds as well. It is not insignificant
that the R\'{e}nyi $q$-entropy enjoys this property for $q\neq1$.
However, its stability fails in the thermodynamic limit. The same
can be said for those parts of the ranges (\ref{pdm1}) and
(\ref{pdp1}) that correspond to negative $s$. For all
positive $q\neq1$ and $s<0$, the unified $(q,s)$-entropy is not stable
in the thermodynamic limit. The case $q=1$ should be separated,
since it includes the Shannon and von Neumann entropies by
definition. From the stability viewpoint, the unified entropies
enjoy desired properties for $1\leq{q}$ and $+1\leq{s}$. This
conclusion is useful, though we do not obtain the complete picture
for all the adopted values of $q$ and $s$.

\section{Subadditivity and triangle inequality}\label{sbtin}

Quantum systems of interest are usually made up of two (or more)
distinct physical systems. Let the state of composite system 'AB'
be described by density operator $\rho_{AB}$ on the product space
$\hh_A\otimes\hh_B$. Then density operators of subsystems 'A' and
'B' are obtained by the partial trace operation \cite{nielsen}.
Namely, we have
\begin{equation}
\rho_A=\tr_B(\rho_{AB})
\ ,\qquad \rho_B=\tr_A(\rho_{AB})
\ , \label{rhoab}
\end{equation}
where the partial traces are taken over $\hh_B$ and $\hh_A$
respectively. Hence we are interested in how used entropic and
other measures may be changed by the partial trace operation.
Except for states of a kind $\rho_{AB}=\rho_{A}\otimes\rho_{B}$,
the question is sufficiently difficult to resolve. In the paper
\cite{lidar1}, this problem is discussed for those unitarily
invariant norms that are multiplicative over tensor products. Some
results have also been obtained for Uhlmann's partial fidelities
\cite{rast092} and the Ky Fan norms \cite{rast10}. For
$k=1,\ldots,d$, the Ky Fan $k$-norm of operator $\ax$ is defined
by \cite{bhatia97}
\begin{equation}
\|\ax\|_{(k)}:=\sum\nolimits_{j=1}^{k}\sigma_j^{\downarrow}(\ax)
\ , \label{kfndf}
\end{equation}
where the singular values should be put in the decreasing order.
The Ky Fan norms and the Schatten norms form especially important
classes of unitarily invariant norms. A unitarily invariant norm
$|||\centerdot|||$ is a norm that satisfies
$|||\ax|||=|||\um\ax\vm|||$ for each $\ax$ and all unitary $\um$,
$\vm$ \cite{bhatia97}.

In the classical regime, the Shannon entropy enjoys the
subadditivity property. The von Neumann entropy is also
subadditive (see, e.g., section 11.3.4 in \cite{nielsen}), i.e.
\begin{equation}
{\rm{S}}(\rho_{AB})\leq{\rm{S}}(\rho_{A})+{\rm{S}}(\rho_{B})
\ . \label{vnsub}
\end{equation}
The subadditivity property was generally assumed to be true for
the Wigner--Yanase entropy \cite{luo07}, until Hansen provided a
counterexample \cite{hansen07}. Another counterexample was given
in \cite{seiringer07}. Meantime, if the bipartite state is
pure then it is sufficient for the subadditivity. Other sufficient
conditions for subadditivity of the Wigner--Yanase entropy are
obtained in \cite{CH10}. It is easy to verify that the Shannon
entropy enjoys the strong subadditivity. The quantum analog holds
for the von Neumann entropy, but the proof turns out to be
hard enough (see, e.g., the recent review \cite{JR10} and
references therein). In the classical regime, the Tsallis
$q$-entropy enjoys even the strong subadditivity for $q>1$
\cite{sf06}. In the finite-dimensional quantum case, we have the
subadditivity for $q>1$:
\begin{equation}
{\rm{H}}_q(\rho_{AB})\leq{\rm{H}}_q(\rho_{A})+{\rm{H}}_q(\rho_{B})
\ . \label{qtsub}
\end{equation}
This result has been conjectured by Raggio \cite{raggio} and later
proved by Audenaert \cite{auden07}. It turns out that the reasons
of the paper \cite{auden07} can be extended to many of the quantum
unified entropies. The key point is an inequality with the Schatten
$q$-norms of the reduced densities (\ref{rhoab}), namely (see theorem 1 in \cite{auden07})
\begin{equation}
\|\rho_{A}\|_{q}+\|\rho_{B}\|_{q}\leq1+\|\rho_{AB}\|_{q}
\ . \label{sqab}
\end{equation}
Due to this result, we obtain the following statement.

\begin{Thm}\label{sbad}
For $q>1$ and $s\geq{q}^{-1}$, the quantum unified entropy is
subadditive, that is
\begin{equation}
\rqm(\rho_{AB})\leq\rqm(\rho_{A})+\rqm(\rho_{B})
\ . \label{unsub}
\end{equation}
\end{Thm}

{\bf Proof.} Following the paper \cite{auden07}, we introduce the
two $2$-dimensional vectors
\begin{equation}
\xp=
\begin{pmatrix}
 \|\rho_{A}\|_{q} \\
 \|\rho_{B}\|_{q}
\end{pmatrix}
\ , \qquad
\zp=
\begin{pmatrix}
 1 \\
 \|\rho_{AB}\|_{q}
\end{pmatrix}
\ . \label{xyvdf}
\end{equation}
Consider the Ky Fan norms of these 2-vectors. First, we have
$\max\bigl\{\|\rho_{A}\|_{q},\|\rho_{B}\|_{q}\bigr\}\leq1$,
since $q$-norms of any density operator with trace one do not exceed
one. Second, the formula (\ref{sqab}) takes place. For all
$k=1,2$, we then write
\begin{equation}
\|\xp\|_{(k)}\leq\|\zp\|_{(k)}
\ . \label{xsmy}
\end{equation}
In other words, the vector $\xp$ is weakly submajorized by $\zp$.
By Ky Fan's dominance theorem (see, e.g., theorem IV.2.2 in the
book \cite{bhatia97}), we have $|||\xp|||\leq|||\zp|||$ for each
unitarily invariant norm. In particular, there holds
$\|\xp\|_{p}\leq\|\zp\|_{p}$ for all the vector $p$-norms. The
last relation can be rewritten as
\begin{equation}
\bigl[\tr(\rho_{A}^{{\,}q})\bigr]^{p/q}+\bigl[\tr(\rho_{B}^{{\,}q})\bigr]^{p/q}
\leq1+\bigl[\tr(\rho_{AB}^{{\,}q})\bigr]^{p/q}
\ . \label{tabpq}
\end{equation}
Putting $s=p/q$ and using the definition (\ref{qundef}), we see
that the relation (\ref{tabpq}) is equivalent to (\ref{unsub}).
Because of $p\geq1$, the reasons hold for $s\geq1/q$.
$\blacksquare$

Thus, we have established the subadditivity of the quantum unified
entropy in a wide range of parameter values. Namely, the entropy
of a state of the whole system is smaller than the sum of the
entropies of its reduced density matrices. In this regard, the
notion of unified entropy concurs with the intuitive reason that a
situation becomes more uncertain when only partial information is
available. For product states of a kind
$\rho_{AB}=\rho_{A}\otimes\rho_{B}$, additive properties have been
examined in the paper \cite{hey06}. It was claimed there that the
subadditivity of the form (\ref{unsub}) fails in the parameter
range
\begin{equation}
\bigl\{(q,s):{\>}1<q,{\>}s<0\bigr\}\cup\bigl\{(q,s):{\>}0<q<1,{\>}0<s\bigr\}
\ . \label{pdmsad}
\end{equation}
We have proved the subadditivity in the range
$\bigl\{(q,s):{\>}1<q,{\ }q^{-1}\leq{s}\bigr\}$. Although we have
incomplete picture of the subadditivity, the question is resolved
for mostly utilized values of the parameters. Like the case of the
von Neumann entropy, the subadditivity inequality (\ref{unsub})
leads to the triangle (or ''Araki-Lieb'') inequality \cite{AL70}.

\begin{Thm}\label{trinq}
For $q>1$ and $s\geq{q}^{-1}$, the quantum unified entropy
satisfies the triangle inequality
\begin{equation}
\bigl|\rqm(\rho_{A})-\rqm(\rho_{B})\bigr|\leq\rqm(\rho_{AB})
\ . \label{arlin}
\end{equation}
\end{Thm}

{\bf Proof.} Let us put the reference system 'C' in such a way
that the triple system 'ABC' is being in a pure state
$|\Psi_{ABC}\rangle$. This vector is a purification of given
density operator $\rho_{AB}$, namely \cite{nielsen}
\begin{equation}
\rho_{AB}=\tr_{C}|\Psi_{ABC}\rangle\langle\Psi_{ABC}|
\ , \qquad
\rho_{C}=\tr_{AB}|\Psi_{ABC}\rangle\langle\Psi_{ABC}|
\ . \label{papuc}
\end{equation}
It follows from the Schmidt decomposition of $|\Psi_{ABC}\rangle$
that the density matrices $\rho_{AB}$ and $\rho_{C}$ have the same
non-zero eigenvalues (see, e.g., section 2.5 in \cite{nielsen}).
Hence $\rqm(\rho_{AB})=\rqm(\rho_{C})$. In a similar manner, we
have $\rqm(\rho_{A})=\rqm(\rho_{BC})$. Combining these two points
with the inequality
\begin{equation}
\rqm(\rho_{BC})\leq\rqm(\rho_{C})+\rqm(\rho_{B})
\label{bcsub}
\end{equation}
finally gives $\rqm(\rho_{A})-\rqm(\rho_{B})\leq\rqm(\rho_{AB})$.
By symmetry, we also have
$\rqm(\rho_{B})-\rqm(\rho_{A})\leq\rqm(\rho_{AB})$. So the claim
(\ref{arlin}) is provided. $\blacksquare$

To sum up, we see the following. For $q>1$ and $s\geq{q}^{-1}$,
the quantum unified entropy enjoys the subadditivity properties
similar to the von Neumann entropy. In particular, the triangle
inequality holds. In general, these results may enough simplify
the use of unified entropies for composite systems. It is
important, because the partial trace operation is usually
inevitable. Note that the above proofs are all restricted to the
discrete case. As was pointed out in \cite{auden07}, some essential
reasons of the result (\ref{sqab}) fail in the case of continuous
distributions. At the same time, finite-dimensional setting is
adequate for many problems of quantum information theory
\cite{nielsen}.

\section{Entropy and projective measurements}\label{proen}

One of physically important properties of the von Neumann entropy
is related to the measurement process \cite{nielsen}. A quantum
measurement is described by a set $\{\mm_i\}$ of measurement
operators. Let $\rho$ be density operator of the quantum system
right before the measurement. Separate terms of the sum
\begin{equation}
\sum\nolimits_i \mm_i{\,}\rho{\,}\mm_i^{\dagger}
\label{mrom}
\end{equation}
are related to different outcomes of the measurement. Since the
probability of $i$th outcome is equal to
$\tr\bigl(\mm_i^{\dagger}\mm_i\rho\bigr)$, the completeness
relation
\begin{equation}
\sum\nolimits_i \mm_i^{\dagger}\mm_i=\pen
\label{grid}
\end{equation}
must be enjoyed \cite{nielsen}. So the operators
$\mm_i^{\dagger}\mm_i$ form a generalized resolution of the
identity $\pen$ (or a ''positive operator-valued measure''). A
standard projective measurement is described by the set
$\{\np_j\}$ of mutually orthogonal projectors which form an
orthogonal resolution of the identity. As a rule, projective
measurements are easier to realize experimentally. In effect,
projective measurements cannot decrease the von Neumann entropy
(see, e.g., theorem 11.9 in \cite{nielsen}). In opposite,
generalized measurements can decrease the entropy. We shall
consider this issue for the quantum unified entropy. For given
projective measurement $\{\np_j\}$, the output density operator is
expressed as \cite{nielsen}
\begin{equation}
\wrho=\sum\nolimits_j \np_j{\,}\rho{\,}\np_j
\ . \label{afms}
\end{equation}
In matrix analysis, an operation of such a kind is usually
referred to as ''pinching'' (see, e.g., section IV.2 in the book
\cite{bhatia97}). For given orthogonal resolution $\{\np_j\}$
and any operator $\ax$ on $\hh$, we define its pinching
\begin{equation}
\cp(\ax)=\sum\nolimits_j \np_j{\,}\ax{\,}\np_j
\ . \label{pindef}
\end{equation}
We are interested in the following question. What is a relation
between the traces $\tr({\wrho}^{{\,}q})$ and $\tr(\rho^q)$? If
${\rm{rank}}(\np_j)=1$ for all $j$, then the answer is simply
formulated as
\begin{equation}
\tr({\wrho}^{{\,}q})
{\ }\left\{
\begin{array}{cc}
\geq, & 0<q<1 \\
\leq, & 1<q
\end{array}
\right\}
{\ }\tr(\rho^{q})
\ . \label{orans}
\end{equation}
The claim follows from Jensen's inequality and the known fact that
\begin{equation}
f\bigl(\langle{j}|\ax|j\rangle\bigr)
\leq\langle{j}|f(\ax)|j\rangle
\label{haps}
\end{equation}
with $\ax\in\lsa(\hh)$, convex function $f(x)$ and unit vector
$|j\rangle$. The question is not so obvious, when
${\rm{rank}}(\np_j)\neq1$ in general. For $q\geq1$, the answer
directly follows from the pinching inequality. This useful
inequality states that \cite{bhatia97}
\begin{equation}
|||\cp(\ax)|||\leq|||\ax|||
\label{pininq}
\end{equation}
for each unitarily invariant norm $|||\centerdot|||$. Using the
Schatten $q$-norm ($q\geq1$), we then obtain
$\|\wrho\|_{q}\leq\|\rho\|_{q}$, whence
\begin{equation}
\tr({\wrho}^{{\,}q})\leq\tr({\rho}^{q})
\label{tqg1}
\end{equation}
due to the positivity of eigenvalues. For $0<q<1$, the traces of
interest do not related with unitarily invariant norms. Here some
properties of operator convex functions are required.

Let $x\mapsto{f}(x)$ be a function of real variable. We say $f$ is
''operator convex'' if
\begin{equation}
f\bigl(\theta\ax+(1-\theta)\ay)\bigr)\leq
\theta f(\ax)+(1-\theta)f(\ay)
\label{conop}
\end{equation}
for all $\ax,\ay\in\lsa(\hh)$ and for $0\leq\theta\leq1$
\cite{bhatia97}. The operator concavity implies the inequality with
reversed sign. It is assumed that eigenvalues of a Hermitian
operator lie in some interval $I$ of the real axis. We take
$I=[0;+\infty)$ for positive semidefinite operators. For $0<q<1$,
we shall use the following statement.

\begin{Lem}\label{paqq}
Let $I$ be an interval containing $0$, and let $f$ be an operator
convex function on $I$ such that $f(0)\leq0$. For each operator
$\ax\in\lsa(\hh)$ with ${\rm{spec}}(\ax)\subset{I}$, there holds
\begin{equation}
\tr\bigl[f\bigl(\cp(\ax)\bigl)\bigr]\leq
\tr\bigl[f(\ax)\bigr]
\ . \label{spah}
\end{equation}
\end{Lem}

{\bf Proof.} Various characterizations of operator convex
functions can be found in chapter V of the book \cite{bhatia97}.
In particular (see theorem V.2.3 therein), for all projections
$\np$ and Hermitian operators $\ax$ with spectrum in $I$ we have
\begin{equation}
f\bigl(\np{\,}\ax{\,}\np\bigr)\leq\np{f}(\ax){\,}\np
\ , \label{fap}
\end{equation}
under the precondition on $f$. All the operators
$\np_j{\,}\ax{\,}\np_j$ have mutually orthogonal supports, whence
\begin{equation}
f\bigl(\cp(\ax)\bigl)=\sum\nolimits_j f\bigl(\np_j{\,}\ax{\,}\np_j\bigr)\leq
\sum\nolimits_j \np_j{\,}f(\ax){\,}\np_j
\ . \label{fap1}
\end{equation}
Since any positive operator has the positive trace, we finally
write
\begin{equation}
\tr\bigl[f\bigl(\cp(\ax)\bigl)\bigr]\leq\tr{\!}\left[\sum\nolimits_j\np_j{\,}f(\ax)\right]=\tr\bigl[f(\ax)\bigr]
\ . \label{fap2}
\end{equation}
Here we used the linear and cyclic properties of the trace,
$\np_j^2=\np_j$ and the completeness relation $\sum_j\np_j=\pen$.
$\blacksquare$

Recall the two known results. The function $x\mapsto{x}^q$ is
operator concave on $\lsp(\hh)$ for $0\leq{q}\leq1$ and operator
convex on $\lsp(\hh)$ for $1\leq{q}\leq2$. For instance, these
results are respectively proved as theorem 4.2.3 and theorem 1.5.8
in \cite{bhatia07}. Combining these points with the statement of
Lemma \ref{paqq}, we obtain
\begin{equation}
\tr({\wrho}^{{\,}q})\geq\tr({\rho}^{q})
\label{tqg0}
\end{equation}
for $0<q<1$ and the inequality (\ref{tqg1}) for $1<q\leq2$ solely.
The main result of this section is formulated as follows.

\begin{Thm}\label{ndprm}
Let $\{\np_j\}$ be an orthogonal resolution of the identity and
let the density operator $\wrho$ be defined by the formula
(\ref{afms}). For $q>0$ and all real $s$, there holds
\begin{equation}
\rqm(\wrho)\geq\rqm(\rho)
\ . \label{rqwr}
\end{equation}
\end{Thm}

{\bf Proof.} We assume $q\neq1$, since the case of von Neumann
entropy is well known. The function $y\mapsto y^s/s$ monotonically
increases for $s\neq0$. Combining this with the inequalities
(\ref{tqg0}) and (\ref{tqg1}) gives
\begin{equation}
\frac{1}{s}{\>}\bigl[\tr({\wrho}^{{\,}q})\bigr]^{s}
{\ }\left\{
\begin{array}{cc}
\geq, & 0<q<1 \\
\leq, & 1<q
\end{array}
\right\}
{\ }\frac{1}{s}{\>}\bigl[\tr(\rho^{q})\bigr]^{s}
\ . \label{rons2}
\end{equation}
Since the factor $(1-q)^{-1}$ is positive for $q<1$ and negative
for $q>1$, the relation (\ref{rons2}) is reduced to
\begin{equation}
\frac{1}{(1-q){\,}s}{\>}\bigl[\tr({\wrho}^{{\,}q})\bigr]^{s}
\geq\frac{1}{(1-q){\,}s}{\>}\bigl[\tr(\rho^{q})\bigr]^{s}
\ . \label{rons3}
\end{equation}
By the definition (\ref{qundef}), the relation (\ref{rons3})
implies (\ref{rqwr}) for $q>0$ and all $s\neq0$. When $s=0$, the
quantum R\'{e}nyi entropy (\ref{qredf}) is dealt. Here we merely
rewrite the relations (\ref{rons2}) and (\ref{rons3}) with the
function $y\mapsto\ln{y}$, which is increasing as well. So similar
reasons provide the inequality (\ref{rqwr}) for this case.
$\blacksquare$

Thus, we have proved non-decreasing of the quantum entropy
(\ref{qundef}) under the action of arbitrary projective
measurements for all the considered values of parameters. It is
easy to see that generalized measurements can decrease the quantum
unified entropy. For the von Neumann entropy of a qubit, an
example is presented in exercise 11.15 of the book \cite{nielsen}.
This measurement is such that $\wrho_{00}=\rho_{00}+\rho_{11}$,
$\wrho_{11}=0$. If the state $\rho$ is both impure and diagonal in
the measurement basis, then
\begin{equation}
\tr({\wrho}^{{\,}q})=\wrho_{00}^{{\>}q}
{\ }\left\{
\begin{array}{cc}
<, & 0<q<1 \\
>, & 1<q
\end{array}
\right\}
{\ }\tr(\rho^{q})=\rho_{00}^{{\>}q}+\rho_{11}^{{\>}q}
\ , \label{exans}
\end{equation}
whence $\rqm(\wrho)<\rqm(\rho)$. For $q>0$ and all real $s$, the
quantum unified $(q,s)$-entropy can be decreased by a generalized
measurement. In this regard, the unified entropies are similar to
the von Neumann entropy.

\section{Conclusions}\label{cocl}

We have considered those properties of the unified entropies that
are of primary importance in physical applications. The unified
entropies have originally been proposed as a two-parameter
extension of the Shannon and von Neumann entropies. It is not
surprising that properties are dependent on the parameter values.
Raised mathematical problems are also of some interest in their
own rights. Some bounds in terms of quantum ensembles are derived.
For certain ranges of parameters, the upper continuity bounds on
the quantum unified entropy are established. The derived
inequalities of Fannes type have lead to the stability property in
the finite-dimensional case. Taking the thermodynamic limit, we
have observed the parameter ranges corresponding to both the
stability and its violation. For a wide range of parameters, the
subadditivity and the triangle inequality are also proved. It
turned out that all the unified entropies are non-decreasing under
projective measurements. Overall, we have observed that the
quantum unified $(q,s)$-entropy enjoys most of the desired
properties for $q\geq1$ and $s\geq1$. In particular, negative
values of the parameter $s$ leads to a stability violation, at
least in the thermodynamic limit. The presented discussion can
herewith be viewed as a supplement and development of the
previously given facts about the unified entropies. The obtained
results are also interesting as applications for some
power methods of matrix analysis.

\acknowledgments

The present author is grateful to an anonymous referee for useful
comments.

\end{document}